\documentclass[preprint2]{aastex}

\shorttitle{Dust in WR112}
\shortauthors{Marchenko et al.}

\begin{document}

\title{Massive Binary WR112 and Properties of Wolf-Rayet Dust}


\author{S.V. Marchenko\altaffilmark{1}, A.F.J. Moffat\altaffilmark{1},
W.D. Vacca\altaffilmark{2}, S. C\^ot\'e\altaffilmark{3}}

\and

\author{R. Doyon\altaffilmark{1}}

\altaffiltext{1}{D\'epartement de physique, Universit\'e de Montr\'eal,
C.P. 6128, Succ. Centre-Ville, Montr\'eal, QC, H3C 3J7, Canada, and
Observatoire du mont M\'egantic}
\email{sergey, moffat, doyon@astro.umontreal.ca}

\altaffiltext{2}{Max-Plank-Institut f\"ur extraterrestrische Physik, Postfach 1312, D-85741 Garching, Germany}
\email{vacca@mpe.mpg.de}

\altaffiltext{3}{Canadian Gemini Office, Herzberg Institute of Astrophysics, National Research Council of Canada, 5071 West Sannich Road, Victoria, B.C., V9E 2E7, Canada}
\email{Stephanie.Cote@hia.nrc.ca}

\begin{abstract} 
Some hot, massive, population-I Wolf-Rayet (WR) stars of the carbon subclass are 
known to be  prolific dust-producers. How dust can form in such a hostile 
environment remains a mystery. Here we report the discovery of a relatively
cool, extended, multi-arc dust envelope around the star WR112, most likely formed 
by wind-wind collision in a long-period binary system.  We derive the binary orbital
parameters, the dust temperature and the dust mass distributions in the envelope. We find 
that amorphous carbon is a
main constituent of the dust, in agreement with earlier estimates
and theoretical predictions. However, the characteristic size of the dust grains is estimated 
to be $\sim 1 \mu m$, significantly larger than theoretical limits. 
The dust production rate is $6.1\times10^{-7} M_\odot {\rm yr^{-1}}$
and the total detectable dust mass is found to be about $2.8 \times 10^{-5} M_\odot$ (for $d=4.15$ kpc).
We also show that, despite 
the hostile environment, at least $\sim$20\% of the initially-formed dust may reach the 
interstellar medium. 
\end{abstract}


\section{Introduction}
Classes of  objects known to produce significant quantities of dust include asymptotic giant-branch  
stars, red giants, novae, supernovae (each of which contributes about equal rates, $\sim 10^{-3} \, 
M_\odot {\rm yr}^{-1}$, of dust in the whole Galaxy), along with planetary nebulae and proto-stars (each of 
which yields $\sim$10 times less; Dwek 1985). Surprisingly, some hot, massive, population I WR stars 
also form dust \citep{wdh87}. WR stars are the evolved descendants of massive O-type stars. They consist 
mainly of He-burning cores surrounded by hot envelopes that drive fast, dense winds with average 
mass-loss rates  $\sim 10^{-5} \, M_\odot {\rm yr}^{-1}$ and terminal velocities $v_\infty \sim 1000 - 4000 \, 
km s^{-1}$. There are three successive WR phases, WN, WC and WO, characterized by the dominant emission 
lines of N, C, and O, respectively, in their optical spectra. All dust-making WR stars belong to the 
carbon-rich, hydrogen-poor WC subclass \citep{wil95}. The WR dust-makers are remarkable for two main reasons: (1) 
the absolute rate of formation is very high, up to $10^{-6} \, M_\odot {\rm yr}^{-1}$ in dust alone, and 
occurring in either a periodic (due to enhanced wind-wind compression at or near periastron passage 
in long-period WC + O binaries with eccentirc orbits) or sustained fashion (in single late-type WC 
stars or moderately short-period WC + O binaries with circular orbits); and (2) the dust is formed 
in a hot, extremely hostile environment, where the formation process is  still unknown.

Here we report on near and mid-infrared imaging observations of the dust envelope surrounding the star 
WR 112 (spectral class WC9). The morphology of this envelope provides clues to the nature of the
stellar system, while the photometry allows us to estimate the properties of the dust and the total
dust mass. 

\section{Observations}

WR112 belongs to the group of  5 WC stars with the densest known dust envelopes \citep{vdh96}. We 
observed WR112 along with WR104 and WR118 with the University of Florida mid-IR imager "OSCIR" at 
the Gemini-North 8m telescope using the medium-band 7.9$\mu m$, 12.5$\mu m$ and 18.2$\mu m$ filters 
on 7 May, 2001. For WR112 we supplemented the mid-IR data with near-IR narrow-band $h$ and $k$ images 
taken at the CFHT on 27 June 1999 with the adaptive optics bonnette and KIR camera, and broad-band L$'$ 
images obtained at the IRTF on 28 May 2000 in the `movie burst' ($2\times 8192$  0.01-sec 
exposures) mode of the NSFCAM imager. We also acquired images of point sources (PSF reference stars), 
usually immediately before and after the principal target, for image restoration. The original data 
cubes were split into individual frames, quickly checked and selected for image quality using basic 
image statistics for the OSCIR and CFHT data (initially compensated for atmospheric turbulence), or 
carefully selected using a specifically designed statistical procedure for the NSFCAM images 
(taken without compensation). The individual images were then appropriately shifted to form a final 
image for each wavelength/target/instrument. The final images were 
then restored using the images of the PSF reference stars and a maximum entropy algorithm 
(the  `stsdas.analysis.restore.mem' task of 
IRAF\footnote{IRAF is
distributed by the National Optical Astronomy Observatories, operated by
the Association of Universities for Research in Astronomy, Inc., under
cooperative agreement with the National Science Foundation.} for the CFHT data, enabling us
to reach \slantfrac{1}{2} pixel ($0.018''$) resolution, or the  `stsdas.analysis.restore.lucy' 
task of IRAF for the OSCIR and NSFCAM images, thus reaching 1-pixel resolution of $0.089''$ and 
$0.055''$, respectively).

\section{Dust properties}
While WR104 and WR118 show only modest extensions on scales of several 0.1$''$, 
combination of the original mid-IR images of WR112 reveals three sets of regularly 
spaced arc-like structures extending out $3''$ from the central star (Fig. 1a). 
It is apparent that the dust temperature gradually falls off outwards. The central 
source and two closest arcs form a W-like configuration, the middle part of which 
is also seen in the near-IR (Fig. 1b). The near-IR size of the central source, 
FWHM=$0.089''\pm 0.004''$, is consistent with earlier measurements \citep{rag98,mon02}: 
FWHM=$0.059''-0.073''$. The near-IR SW elongation probably corresponds to the 
previously detected \citep{rag98} small-scale asymmetry in the dust envelope.

WR112 is the fourth case in which dust around a WR star has been mapped in some detail, 
with all the previous examples occurring in WC+O binaries \citep{mar99,mon99,tut99}.
Note that 6 other WR stars have been shown to posess an extended (presumably spherically-symmetric)
envelope \citep{yud01,mon02}. 
WR112 is also known to be a variable non-thermal radio emission source 
\citep{cha99,mon02}, which might indicate  binarity \citep{dou00}. There are 
few viable scenarios for dust production in WR112. For example, the primordial 
circumstellar envelope would not have survived in the hard UV radiation-field 
of the WR progenitor. Periodic outbursts driven by stellar oscillations cannot 
explain the observed dust envelope-shape without invoking a factor of two azimuthal 
variation in the terminal-wind velocity along with an extreme latitudinal 
dependence of the mass-loss. Therefore, we assume that the dust is formed 
in the wind-wind collision zone of a
massive binary system (thus resembling a collimated but rotating `beam' 
for a distant observer) and expelled radially outward at a constant, 
terminal-wind velocity $v_\infty =1200\, km\, s^{-1}$ \citep{roc95}. We then  
calculate the resulting locus of maximum dust density and fit it to the 
observed dust distribution in the restored mid-IR images, taking the necessary 
orbital elements, $P,e,i,\omega , \Omega$ as free parameters. We find a family of 
minimum-$\chi ^2$ 
solutions for counter-clockwise orbits, in which coupling of $e$-$i$ and 
$\omega$-$\Omega$ leads to some degree of degeneracy. Instead of relying solely on the single formal
best fit, we therefore construct a synthetic solution by averaging all the possible 
combinations of solutions for which the $\chi ^2$ values lie within 5\% 
of the absolute minimum. This average solution has period $P=24.8\pm 1.5 \, {\rm yr}$, 
eccentricity $e=0.11\pm 0.11$, orbital 
inclination $i=38.0^{\arcdeg} \pm 3.8^{\arcdeg}$,  periastron angle 
$\omega =269.6^{\arcdeg} \pm 18.5^{\arcdeg}$ and orbital orientation 
$\Omega=33.5^{\arcdeg} \pm 7.9^{\arcdeg}$ (1-$\sigma$ errors). 
In Fig. 2 we plot the solution with the 
highest possible, but still plausible, eccentricity: $e=0.40$. This particular 
model yields a dust locus that is actually very close to that derived 
from the synthetic solution. In general, the model provides fairly good fits to the SW arms, 
although it fails to explain the pronounced NE excursions, as well as the bright SW
protuberance. The latter extends in the direction toward a stellar visual companion $\sim 1''$ 
from WR112 (Wallace, Moffat \& Shara 2002, and our Fig.2), which is not seen at near-IR, and is 
therefore probably also invisible at mid-IR wavelengths. Clearly the dust-production 
rate depends on orbital phase, diminishing as expected around apastron but 
surprisingly also absent around periastron, contrary to the situation in the eccentric, long-period 
colliding-wind WR+O binaries WR137 and WR140 \citep{vdh02,mar99}. The average thickness, 
$\delta r$, of the dust arcs in the deconvolved images relative to the separation between 
the successive arcs, $\Delta r$ (measured at the same orbital phase), gives an estimate 
of the shock-cone opening angle: 
$\Theta \sim 2\pi \times \delta r / \Delta r \sim 60^{\arcdeg} -110^{\arcdeg} $, similar 
to that found in other colliding-wind WR+O systems \citep{mar97,bar01}. In Table 1 we 
list the calculated orbital elements along with the main characteristics of the dust envelope
and the assumed parameters.

To derive the basic dust properties, which fortunately do not depend 
sensitively on the adopted orbital parameters, we first deproject all apparent 
distances by applying  the synthetic average orbit-solution. We calibrate the 
restored images using all available IR-flux measurements of WR112 \citep{wil87,vdh96}, 
assuming that the system has not experienced any large-amplitude IR outbursts. 
In the mid-IR range, the relatively large apertures used in previous ground- and 
space-based observations will tend to smooth out the variability by averaging over 
the output of many dust-formation cycles. We deredden the near-IR data by applying 
a Galactic interstellar-extinction model \citep{are92}, then calculate the net dust fluxes 
after subtracting the stellar component, approximated as 
$F_\lambda \sim \lambda^{-\alpha}$, with $\alpha =2.97$ \citep{mor93}. Then we use 
the relation \citep{hil83}:
\begin{equation}
M_d(r) = {{4\rho F(\lambda , r)d^2 a}\over{ 3 B(\lambda ,T(r)) Q_e (\lambda ,a)}},       
\end{equation}
where  $M_d$ is the dust mass, $F(\lambda , r)$ is the measured flux at a given position $r$, 
$\rho$ is the grain density, $d$=4.15 kpc is the distance to WR112 \citep{vdh01}, $a$ is the grain size, 
$B(\lambda ,T(r))$ is the black-body emissivity, and $Q_e(\lambda ,a)$ 
is the grain emission coefficient. We adopt the general relation 
$Q_e(\lambda , a)/a=k \lambda ^{-s}$  
along with a $T=T_0 r^{-\beta}$ temperature profile \citep{wil87} and measure the 
fluxes $F(\lambda , r)$  
over regions of fixed $7\times 7$ pixel size ($0.623''\times 0.623''$), 
assigning 15\% /30\% errors to $F(\lambda ,r)$ in the inner/outer-most arcs. Taking the ratios of 
fluxes at every given position in the envelope (except the very central part, where 
the spectral distribution cannot be represented by a single temperature) in order to 
reduce the number of free parameters and to be impervious to large positional fluctuations 
in $F(\lambda , r)$, we obtain via $\chi ^2 $-minimization the following estimates:  
$s= 0.90^{+0.22} _{-0.21}$, $T_0 = 320^{+12} _{-13}$ K, $\beta = 0.40 ^{+0.05} _{-0.05}$ 
($T_0$ at $1''$; 95\% confidence intervals). The dust emissivity index $s$, though loosely constrained, 
is close to the range expected for amorphous carbon in the near-mid-IR domain \citep{bor85,suh00}. On the other hand, 
the absence of the telltale 9.7$\mu m$ 
silicate emission feature in the WR112 spectrum \citep{vdh96} rules out the presence 
of any silicate-based mixture. The derived temperature profile follows the $\beta = 0.4$ 
dependence expected in the case of thermal equilibrium. Adopting $\rho =2.0 \, g \, cm^{-3}$ 
for the amorphous carbon dust, we estimate 
the dust mass separately for each $7\times 7$ pixel$^2$ area of each of 
the 3 mid-IR images, then average the results (Fig. 3). We find a total $M_d=5.5\times 10^{28}\, g$, integrated 
over the observed envelope. To compare our result with an earlier theoretical estimate of the mass we (a) 
account for the difference in the adopted distances to WR112 and (b) assume that the observed dust is 
created for 15 yrs (see Fig. 2) during each 25-yr orbital cycle, with 3 complete 
cycles accounted for. A model \citep{zub98} with the assumption of a spherically-symmetric 
envelope gives $\dot M_d=3.8\times 10^{-7} \, M_\odot \, {\rm yr}^{-1}$ (converted to d=4.15 kpc), while our estimate is  
$\dot M_d=6.1\times 10^{-7} \, M_\odot \, {\rm yr}^{-1}$.  Calculation 
of the {\it {integrated}} mass-ratios, 
M(innermost region+arcs) : M(middle arcs) : M(outer-most arcs)=1.0 : 0.6 : 0.2, shows that the dust is 
gradually destroyed while leaving the system.  Both the intense UV radiation field and 
sputtering caused by the high drift-velocities of the dust grains \citep{zub98} may be responsible for 
the dust destruction. Nevertheless, the flattening of the trend  
at large radii in Fig. 3 suggests that some dust does reach the ISM.

Finally, we use eq. 1 to evaluate the characteristic size of the dust grains, 
by comparing via least-square minimization, the observed fluxes to the prediction 
of a simple model assuming that the envelope is populated by similar-sized 
spherical particles. We calculate the Mie absorption coefficients \citep{boh98}, 
$Q_{abs} (\lambda ,a) = Q_e (\lambda , a)$, 
using the optical constants for amorphous carbon \citep{zub96}. As a second free parameter we 
take the integrated (innermost region+arcs) dust mass, allowing it to follow the general 
trend seen in Fig. 3. This gives an estimate of $M_d$ within 25\% of the previously calculated value, 
thus providing a good consistency check. We find a characteristic grain-size $a=0.49 ^{+0.11} _{-0.11} \mu m$ 
(95\% confidence interval), in line with the estimates obtained for WR112 from ISO 
spectroscopy, $a \sim 1 \mu m$ \citep{cha01}. The large grain size poses a serious problem for 
the theory of grain growth, which calls for $a \sim 0.01 \mu m $ \citep{zub98}.

\section{Conclusions}
The recently acquired high-resolution, high signal-to-noise near-mid-IR
images of the WR star WR112 have enabled us to resolve and, for the
first time, study in detail the extended dust envelope thought to be formed 
in the wind-wind collision zone of the long-period, $P \approx 25 \, {\rm yr}$, binary. A simple
approach involving a minimum of assumptions has allowed us to derive the
basic dust properties: We find that the dust emissivity approximately follows the 
trend expected for amorphous carbon.
The radial temperature-distribution shows the dust envelope to be in thermal equilibrium. 
The dust production-rate corresponds to $\sim 6\% $ 
of the total mass-loss rate, if for the latter we assume a typical value of $10^{-5} \, M_\odot \, 
{\rm yr}^{-1}$. 
The characteristic size of the dust particles, $a \sim 1\mu m$, turns out to be much larger 
than expected from state-of-the-art  models of dust growth. We also show that, despite the harsh conditions,
$\sim 20\%$ of the newly formed dust can survive and escape the system, thus enriching
the interstellar medium. This may have a direct implication for the
process of heavy-element enrichment in the early Universe, when massive
stars completely dominated the stellar scene.

\acknowledgments
We thank V. Zubko and P. Williams for helpful discussions, and J.-R. Roy and D. Toomey for assistance in observations. S.V.M., A.F.J.M. and R.D. gratefully acknowledge  financial support from NSERC (Canada) 
and FCAR (Qu\'ebec). This paper is based on observations obtained at the Gemini Observatory, 
which is operated by the Association of Universities for Research in Astronomy, Inc., under 
a cooperative agreement with the NSF on behalf of the Gemini partnership: NSF 
(United States), PPARC (United Kingdom), NRC (Canada), CONICYT (Chile), ARC 
(Australia), CNPq (Brazil) and CONICET (Argentina). The observations were made with 
the mid-infrared camera OSCIR, developed by the University of Florida with support 
from NASA, and operated jointly by the Gemini Observatory and the University of Florida Infrared Astrophysics Group.

\clearpage

\begin{figure}
\caption{~a: False-color image of WR112 produced by log-scaling, normalizing and combining the 7.9, 12.5 and 18.2$\mu m$ images as blue, green and red. The faintest details are $\sim 10^{-3}$ of the central source.  The inset shows a point-spread-function standard at the same scale, used for image restoration.
Fig. 1b: Zoom on the central source: combination of the maximum-entropy deconvolved, log-scaled and normalized images of WR112, with the narrow-band images $h$ ($\lambda _0=1.64\mu m$, FWHM=$0.02\mu m$) and  $k$ ($\lambda _0=2.17\mu m$, FWHM=$0.02\mu m$) as blue and green, respectively, and the broadband L$'$ image ($\lambda _0 = 3.75\mu m$)  as red. The deconvolved and appropriately enlarged  7.9$\mu m$ image is shown in contours. Small black stars mark the same central position on each image. \label{fig1}}
\end{figure}

\clearpage 

\begin{figure}
\caption{ Deconvolved 18.2m image of WR112 with the overplotted model fit corresponding to the high-eccentricity case, $ P=23.5 {\rm yr}$, $e=0.40$, $i=35^{\arcdeg}$, $\Omega=40^{\arcdeg}$, $\omega=255^{\arcdeg}$ (for the {\it {average}} derived parameters see Table 1), with the red line tracing the locus of visible dust, then gray line beyond. The grossly enlarged, projected orbit is shown in the lower left corner. The red dotted line points toward periastron. A small blue cross depicts the position of the visual companion.\label{fig2}}
\end{figure}

\begin{figure}
\caption{The integrated (within $7\times 7$ pixel areas) dust mass in grams. Filled
triangles mark the central part of the image along with the  two innermost arcs.
Filled (open) squares/circles correspond to the SW (NE) middle/outer-most arc. 
The dotted line shows the trend expected in the case of mass conservation. \label{fig3}}
\end{figure}

\clearpage

\begin{deluxetable}{lll}
\tabletypesize{\scriptsize}
\tablecaption{Orbital parameters (`synthetic solution'), basic dust characteristics and 
assumed values. \label{tbl-1}}
\tablewidth{0pt}
\tablehead{
\colhead{Parameter} & \colhead{Value}   & \colhead{Errors\tablenotemark{a}}
}
\startdata
$P$, yr & 24.8  & $\pm 1.5$ \\
$e$     & 0.11  & $\pm 0.11$ \\
$i$     & $38.0^{\arcdeg}$ & $\pm 3.8^{\arcdeg}$ \\
$\omega$ & $269.6^{\arcdeg}$ & $\pm 18.5^{\arcdeg}$ \\
$\Omega$ & $33.5^{\arcdeg}$  & $\pm 7.9^{\arcdeg}$  \\
$s$      &  0.90          & -0.21, +0.22 \\ 
$T_0$\tablenotemark{b}, K & 320 & -13,+12 \\
$\beta$  &  0.40 & -0.05, +0.05 \\
$M_d$, g & $5.5\times 10^{28}$ & --- \\
$\dot M_d$, $ {\rm M_\odot \, yr}^{-1}$ & $6.2\times 10^{-7}$ & ---\\
$a$, $\mu m$ & 0.49 & -0.11, +0.11 \\ 
$v_{\rm {dust}}$, $\rm {km \, s^{-1}}$ & 1200 & assumed \\
$\rho_{\rm {dust}}$, $\rm {g \, cm^{-3}}$ & 2.0 & assumed \\
$d$, $\rm {kpc}$ & 4.15 & assumed \\
\enddata


\tablenotetext{a}{Either 1-$\sigma$ or 95\% confidence interval}
\tablenotetext{b}{$T_0$ at $1''$}

\end{deluxetable}

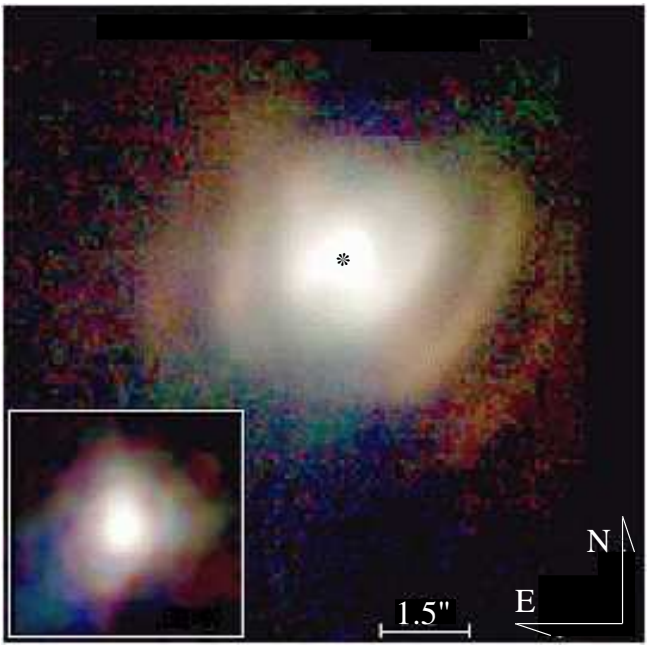
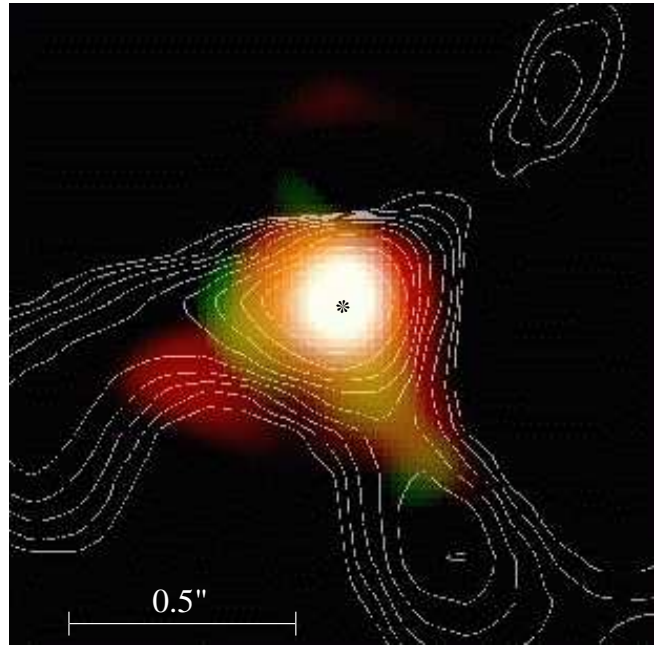
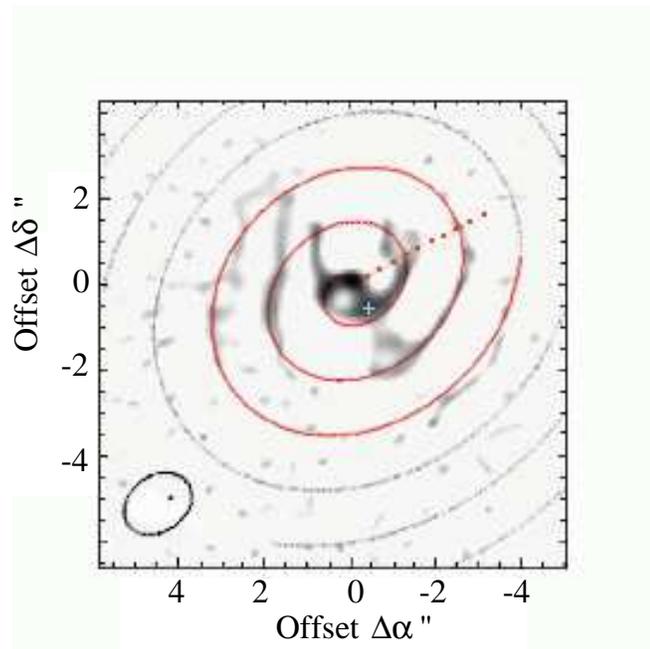
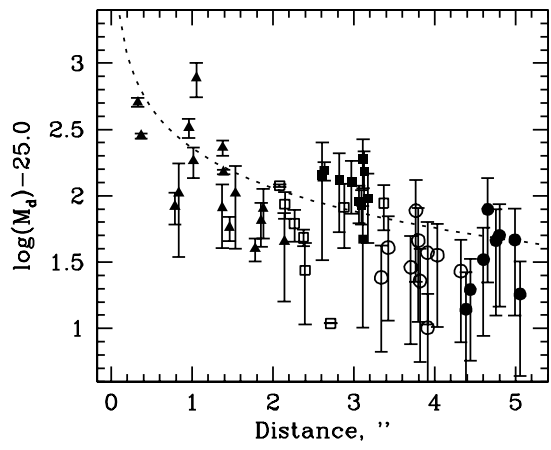
\end{document}